# Conductance-Photoacoustic Spectroscopy for OneSignal Measurement of Multi-components

Ruobin Zhuang, Jianfeng He and Huadan Zheng*

*Department of Optoelectronic Engineering, Jinan University, Guangzhou, 510632, China*
*\*Email correspondance : zhenghuadan@jnu.edu.cn*

Ensuring safety and efficiency in emerging hydrogen–hydrocarbon fuel systems requires accurate measurement of multiple gas components in real time. However, existing detection techniques generally lack the capability to quantitatively measure hydrogen and natural gas constituents simultaneously. Here, we present a novel conductance–photoacoustic spectroscopy (CPAS) method that integrates platinum-modified conductance measurements with beat-frequency photoacoustic detection. By bridging a quartz tuning fork with a platinum microwire, our approach enables direct monitoring of hydrogen concentration via frequency modulation, while simultaneously capturing propane's photoacoustic signal with a single detection channel. Experimental results confirm that the platinum microwire effectively fine-tunes the tuning fork's mechanical properties for high-sensitivity hydrogen measurement, and the beat-frequency photoacoustic signals from propane absorption reveal complementary hydrocarbon concentration information. This unified sensor design is inherently compact, rapid, and calibration-free, making it particularly suitable for applications that demand real-time multiparameter gas analysis, including industrial process control and environmental monitoring. Taken together, these findings demonstrate that combining conductance and photoacoustic spectroscopies into a single integrated platform significantly advances the state of multi-component gas detection and holds promise for further enhancements in sensitivity and adaptability.

Simultaneous detection of natural gas constituents, such as methane, ethane, and propane, alongside hydrogen, is crucial for ensuring both safety and optimal combustion performance in modern energy systems [1], [2]. These mixed-fuel environments can exhibit highly variable explosion pressures and pressure rise rates due to the inclusion of hydrogen, which possesses different flammability characteristics compared with hydrocarbon-based fuels [1]. Moreover, with the growing push toward "power-to-gas" strategies that involve injecting hydrogen into natural gas grids, the composition of fuel gases can fluctuate significantly [2]. Such variations necessitate continuous, real-time monitoring to maintain reliable operation conditions, prevent potential accidents, and ensure compliance with energy standards. Hence, the ability to accurately quantify each component in mixed hydrogen–hydrocarbon streams enables safer handling, facilitates compliance with emission regulations, and supports the development of cleaner, more efficient combustion technologies.

Among various gas-detection methods, conductance spectroscopy (CS) and photoacoustic spectroscopy (PAS) stand out for their high sensitivity, compact design, and suitability for real-time measurements [3], [4]. Conductance spectroscopy typically involves modifying a quartz tuning fork with a micro wire and determining the target analyte concentration by monitoring changes in the overall electrical conductance of the quartz-wire system [3]. Meanwhile, photoacoustic spectroscopy leverages the absorption of laser light by gas molecules to produce acoustic waves that are converted into an electrical signal via the piezoelectric effect, enabling highly selective, sensitive detection in a small footprint [4]. However, as far as we know, all existing approaches still fall short of enabling simultaneous quantification of both carrier and target gases (e.g., hydrogen and hydrocarbon fuels) in a single measurement, underscoring the need for advanced techniques that can capture all critical gas components at once.

Hydrogen can be measured through multiple approaches, including catalytic sensors, electrochemical sensors, semiconductor sensors, and metal–insulator sensors [5], [6]. Among the catalytic sensor materials, platinum has gained particular attention due to its high catalytic activity, chemical stability, and capacity to operate across a wide temperature range with minimal cross-sensitivity to other gases [5]. Additionally, platinum-based sensors have been demonstrated to accurately and reversibly detect hydrogen even at relatively low concentrations, without significant interference from carrier gases like methane or nitrogen [6].

Beat-frequency photoacoustic spectroscopy (BF-PAS) has emerged as a rapidly advancing variant of photoacoustic spectroscopy, exploiting the difference between the laser modulation frequency and the quartz tuning fork (QTF) resonance frequency to generate a beat-frequency signal [7], [8]. This beat signal's frequency can be tracked in real time to continuously calibrate the QTF resonance, while its amplitude correlates with the target gas concentration, thus enabling a calibration-free strategy for trace-gas detection [7]. In recent implementations, BF-PAS has already demonstrated ppb-level sensitivity in applications such as ammonia monitoring [8].

In this work, we propose, for the first time, a conductance–photoacoustic spectroscopy (CPAS) technique that unifies the high hydrogen sensitivity of platinum-based conductance spectroscopy with the real-time, calibration-free benefits of beat-frequency photoacoustic spectroscopy. Specifically, we employ a Pt-modified microwire in conductance mode to accurately capture hydrogen concentration, while the beat-frequency photoacoustic signal simultaneously offers insights into natural gas components—both from a single measurement. Such a fully integrated sensor architecture is inherently compact and yields rapid detection, making it well-suited for industrial and environmental monitoring applications. By combining the complementary strengths of conductance and photoacoustic methods into one platform, this study paves the way for fast, real-time measurements of multi-component gas mixtures without requiring frequent recalibrations.

Figure 1 illustrates the core component of the conductance–photoacoustic spectroscopy (CPAS) platform. A platinum (Pt) microwire of 15um diameter and 2mm length is precisely fabricated and then bridged between the two prongs of the quartz tuning fork (QTF). Due to the inherent ductility of Pt, the microwire provides an elastic modulus that, when coupled with the QTF, leads to a unified resonant system whose overall resonance frequency is effectively tuned by the Pt microwire. On either side of the QTF prongs, a pair of stainless-steel micro acoustic resonators are mounted to sustain the acoustic wave along the laser propagation path. Each resonator has a total length of approximately half of the acoustic wavelength, subject to minor geometric corrections due to non-ideal half-wave resonance. The QTF itself is centered at the position of maximum acoustic pressure within these resonators, with a gap of around 50um between the resonator ends and the QTF surface. This design ensures optimal coupling of the acoustic energy to the QTF while allowing the fork to vibrate freely.

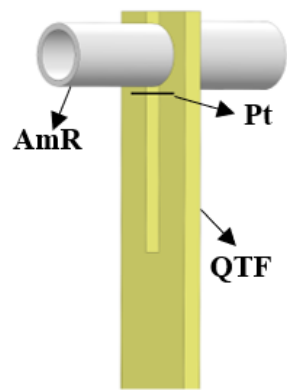

Fig. 1 Spectrophone of Conductance-Photoacoustic Spectroscopy (CPAS).

Figure 2 presents a schematic of the experimental setup used for the CPAS measurements. A function generator produces two signals: a sinusoidal wave for modulating the laser, and a step ramp signal for generating the beat frequency; these two signals are superimposed and sent to the laser driver. The system employs an interband cascade laser (ICL) centered at 3368 nm to measure propane absorption. Although propane is chosen here for demonstration purposes, the same approach can be extended to methane or other hydrocarbons simply by selecting a laser chip tuned to the appropriate absorption line. After passing through a collimator, the laser beam traverses the gas cell containing the mixture of hydrogen, nitrogen, and propane at controlled concentrations, then reaches the core CPAS device. The Pt microwire on the QTF selectively responds to hydrogen by altering the mechanical frequency of the fork, while the acoustic waves generated by absorbing propane produce vibrations that modulate the QTF amplitude. The output from the tuning fork is then pre-amplified and demodulated using a lock-in amplifier, before being digitized and recorded. Notably, with this CPAS system, a single detection signal enables simultaneous determination of both hydrogen and hydrocarbon concentrations.

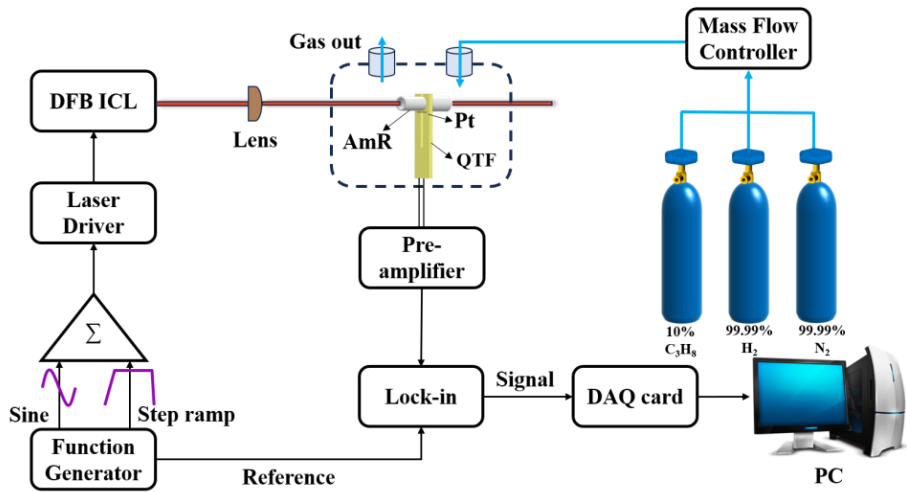

Fig. 2 Experimental setup of CPAS for gas analysis.

Figure 3 illustrates typical CPAS signals acquired under these operating conditions. The pristine QTF resonance frequency is initially 32760.2 Hz. Upon bridging the QTF with the Pt microwire, the fork's effective resonance shifts to 32667.9 Hz, reflecting the influence of the Pt's elasticity. For beat-frequency generation, the sinusoidal modulation is set at 32600 Hz, resulting in a difference of ~70 Hz relative to the QTF's resonance. The amplitude of the modulation signal is maintained at 15 mA. Concurrently, a pulsed harmonic signal sweeps the laser current from 90.1 mA to 92.1mA, encompassing the propane absorption line at 3369.74nm. As shown in Figure 3, the detected beat-frequency signal exhibits an exponential decay. By examining the periodicity (i.e., the frequency) in the horizontal direction (marked in blue), we identify the QTF's effective resonance, which shifts in response to hydrogen concentration. Calibration data then relate this frequency shift to the hydrogen concentration of 25%. Meanwhile, the red markers in the vertical direction track the amplitude of the beat signal peaks, which directly correlates with the propane absorption level. The difference between a peak and its subsequent trough quantifies the signal amplitude change, thereby yielding the propane concentration of 2%. Thus, with a single measurement and a single output trace, the CPAS approach simultaneously determines hydrogen and propane concentrations, showcasing its unique advantage in integrated, dual-species gas sensing.

In summary, this work demonstrates how conductance–photoacoustic spectroscopy unifies the strengths of platinum-based conductance sensing with beat-frequency photoacoustic detection to simultaneously quantify hydrogen and natural gas components. By modifying a quartz tuning fork with a platinum microwire, we achieved real-time tracking of hydrogen-induced frequency shifts, while photoacoustic signals provided concurrent insight into propane concentrations. The key finding is that both measurements can be obtained from a single signal, enabling compact, rapid, and calibration-free analysis of multi-component gas mixtures. This advance is important because it allows for robust gas detection in scenarios where both safety and efficiency are paramount, such as hydrogen-enriched fuel pipelines and combustion processes. The approach outlined here can be extended to other relevant gas mixtures through appropriate laser wavelength selection and sensor design. Hence, this study not only paves the way for more integrated and versatile sensor architectures but also underpins future explorations that could refine the sensitivity, range, and adaptability of the conductance–photoacoustic platform.

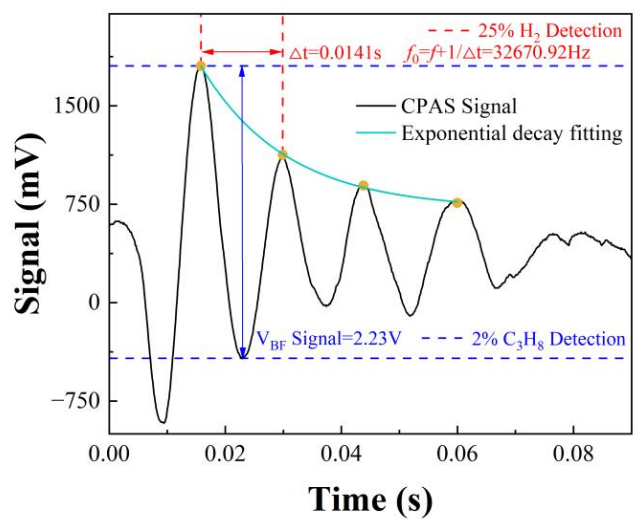

Fig. 3 The signal of CPAS for $H_2$ and $C_3H_8$ mixture.

Conflict of interest

The authors declare that there are no conflicts of interest.

Acknowledgment

This work is supported by the National Natural Science Foundation of China (62475111).